\pdfoutput=1
\documentclass{aa}  

\usepackage{graphicx}
\usepackage{txfonts}
\usepackage{lscape}
\usepackage{booktabs}
\usepackage{threeparttablex}
\usepackage[bookmarks=false,colorlinks=true,linkcolor=black,citecolor=black,filecolor=black,urlcolor=cyan]{hyperref}

\begin{document}

   \title{The dependence of gamma-ray burst X-ray column densities on the model for Galactic hydrogen}

   \author{R. Arcodia
          \inst{1,2}
          \and
          S. Campana\inst{2}
          \and
          R. Salvaterra\inst{3}
          }

   \institute{Department of Physics G. Occhialini, University of Milano-Bicocca, Piazza della Scienza 3, 20126 Milano, Italy\\
              \email{r.arcodia@campus.unimib.it}
         \and
             INAF-Osservatorio Astronomico di Brera, Via Bianchi 46, I-23807 Merate (LC), Italy\\
             \email{sergio.campana@brera.inaf.it}
         \and
	          INAF, IASF Milano, via E. Bassini 15, 20133 Milano, Italy
             }

   \date{Received 18 February 2016; accepted 25 March 2016}

 
  \abstract{
	We study the X-ray absorption of a complete sample of 99 bright \emph{Swift} gamma-ray bursts. Over the last few years, a strong correlation between the intrinsic X-ray absorbing column density ($N_H(z)$) and the redshift was found. This absorption excess in high-$z$ GRBs is now thought to be due to the overlooked contribution of the absorption along the intergalactic medium, by means of both intervening objects and the diffuse warm-hot intergalactic medium along the line of sight. In this work we neglect the absorption along the IGM, because our purpose is to study the eventual effect of a radical change in the Galactic absorption model on the $N_H(z)$ distribution. Therefore, we derive the intrinsic absorbing column densities using two different Galactic absorption models, the Leiden Argentine Bonn $H_I$ survey and the more recent model including molecular hydrogen. We find that, if on the one hand the new Galactic model considerably affects the single column density values, on the other hand there is no drastic change in the distribution as a whole. It becomes clear that the contribution of Galactic column densities alone, no matter how improved, is not sufficient to change the observed general trend and it has to be considered as a second-order correction. The cosmological increase of $N_H(z)$ as a function of redshift persists and, in order to explain the observed distribution, it is necessary to include the contribution of both the diffuse intergalactic medium and the intervening systems along the line of sight of the GRBs.

			}

   \authorrunning{R. Arcodia et al.} 
   \maketitle
   
   \defcitealias{Campana:complete}{C12}
   \defcitealias{Willingale:TotalH}{W13}
  
%

\section{Introduction}
Long Gamma-Ray Bursts (LGRBs) are associated with type Ibc supernova explosions \citep{Woosley:SN} and observed into galactic regions with high star formation rate \citep{Fruchter:environment}: this allows us to infer that LGRBs have massive stars as progenitors. Moreover, their circumburst medium is thought to be denser than the typical star formation regions, as we can see from the high absorbing column density values measured in X-rays \citep{Galama:highNh,Stratta:XRabs,Gendre:XRaft}. Both metals in the circumburst medium and within the whole host galaxy may contribute to the intrinsic column density $N_H(z)$. In addition, the X-ray absorption occurs within our Galaxy (parametrized by $N_H(Gal)$) and along the line of sight due to the intergalactic medium (IGM), by means of a warm-hot diffuse medium pervading it or, mainly, by means of compact discrete intervening systems (galactic halos, galaxy groups) casually placed along the line of sight of the GRB \citep[and references therein]{Behar:IGM,Eitan:Xrab,Starling:evoluzIGM,Campana:missing}.

\citet[hereafter C12]{Campana:complete} worked out a redshift distribution of the intrinsic column density of a complete sample of 58 \emph{Swift} LGRBs, assuming the absorption along the IGM to be negligible. It emerged an increasing trend of $N_H(z)$ with the redshift of the GRB, indicating a lack of non-absorbed GRBs at high redshift or, alternatively, an absorption excess in high-$z$ GRBs. The sample in C12, named BAT6 \citep{Salvaterra:complete}, has been selected above a specific threshold of the gamma peak flux and has a high completeness in redshift, which is fundamental not only to exclude the hypothesis that the $N_H(z)-z$ correlation is due to observative biases, but also to provide the real intrinsic column density values calculated at the GRB redshift, taking account of the scaling cosmological factor $\sim(1+z)^{2.4}$  \citep[see][]{Campana:scaling}. Recently, it has been shown that a model including the (previously overlooked) contributions of both the intervening objects and the diffuse warm-hot intergalactic medium along the line of sight could explain qualitatively and quantitatively the observed $N_H(z)$ distribution (\citealp{Campana:missing}; \citealp[see also][]{Behar:IGM,Eitan:Xrab,Starling:evoluzIGM}).

We now derive the intrinsic column densities still neglecting the contribution along the IGM (then allocating all the absorption in excess to the host galaxy), because our purpose is to study the eventual effect on the $N_H(z)$ distribution of a radical change in the Galactic absorption model. As a matter of fact, we hold the Galactic component fixed, using the Leiden Argentine Bonn (LAB) $H_I$ survey \citep[largely adopted by previous literature, including C12]{Kalberla:LAB} to verify the $N_H(z)-z$ correlation and, in addition, we consider the new Galactic absorption model introduced by \citet[hereafter W13]{Willingale:TotalH}. This new model completes the $H_I$ distribution of the LAB survey, obtained in radio, adding the contribution of the molecular hydrogen, extracted from the IR dust absorption in our Galaxy. Therefore, it reports generally greater values than the LAB survey and, consequently, the new intrinsic column density values are expected to be smaller. Moreover, we derive our results using an extended complete sample of 99 \emph{Swift} LGRBs, named BAT6ext, updated to GRB 140703A with the bursts detected by \emph{Swift} that, since the construction of the BAT6, have satisfied the same selection criteria \citep{Pescalli:BAT6ext}. In this, we enlarge our sample from 58 to 99 LGRBs, but we slightly lose redshift completeness from $95\%$ to $82\%$.
\section{X-ray absorbing column density distribution}
The intrinsic column densities were computed from GRBs X-ray afterglow spectral fits using the \emph{UK Swift Science Data Centre} spectra repository, that provides \emph{Swift}/XRT data suitable for scientific analysis \citep{Evans:UKSSDC}. All the results, listed in Table~\ref{tab:longtable}, were obtained by selecting a specific time interval when there are no strong spectral variations. This is obtained by strictly selecting time intervals when the hardness ratio (between counts in the $1.5-10\,$keV and in the $0.3-1.5\,$keV bands) is constant. For an afterglow spectrum modelled as a power law, this condition assures that no strong spectral changes are occurring. This is particularly important because any curvature estimate of the X-ray spectrum would reverberate into the column density, producing in turn biased values \citep{Butler:HRevolution}.

\begin{figure}[!h]
	\centering
	\includegraphics[width=\columnwidth]{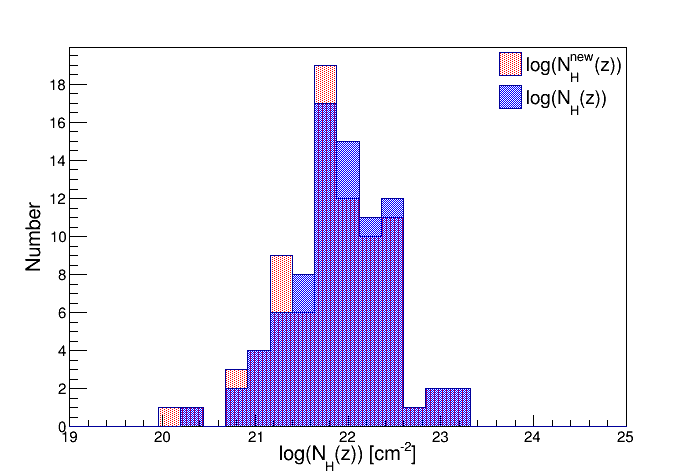}
	\caption{Intrinsic X-ray absorbing column density distributions of the BAT6ext sample. The dotted (red) histogram is related to the $N^{new}_H(z)$ distribution, while the thicker dotted (blue) histogram shows the $N_H(z)$ distribution.}
	\label{fig:histo}
\end{figure}

Each spectrum was then fitted by a power-law, absorbed by both a Galactic component, held fixed, and an intrinsic component at the redshift of the GRB, modelled with \texttt{TBABS} and \texttt{ZTBABS}, respectively, within \texttt{XSPEC} (version 12.9.0). It is important to note that the results in C12 were obtained by using different models, namely \texttt{PHABS} and \texttt{ZPHABS} (the Leicester \emph{Swift} automatic spectral analysis tool adopted the TBabs models only in 2014\footnote{See the change log here \href{http://www.swift.ac.uk/xrt_spectra/docs.php}{www.swift.ac.uk/xrt\_spectra/docs.php}}). Therefore, the intrinsic column density values of the BAT6 GRBs, in common between the two works, are slightly different. Abundances from \citet{Wilms:abund} and cross-sections from \citet{Verner:xsect} were used. Table~\ref{tab:longtable} shows the intrinsic column density values obtained using the Galactic absorption models provided by both LAB and W13 surveys, the latter being labelled as \emph{new}. For the 18 GRBs of the BAT6ext sample having no spectroscopic redshift, we fixed $z$ to be zero and, consequently, the intrinsic column densities have to be considered as lower limits.

The two $N_H(z)$ distributions are showed in Fig.~\ref{fig:histo} and can be described by a Gaussian, with a mean value of $\log (N_H/\rm{cm}^{-2})=21.93\pm0.54$ and $\log (N_H/\rm{cm}^{-2})=21.84\pm0.61$, for the LAB survey and the W13 Galactic model, respectively. The latter mean value is only slightly smaller but fully consistent with the former. This likely indicates that the change in the Galactic column densities plays a minor role in shaping the intrinsic column density distribution. Both values are consistent with the distribution of column densities reported in C12, that is $\log (N_H/\rm{cm}^{-2})=21.7\pm0.5$. This indicates that the impact of different \emph{Swift}/XRT response matrices\footnote{See the release note at \href{http://www.swift.ac.uk/analysis/xrt/files/SWIFT-XRT-CALDB-09_v20.pdf}{http://www.swift.ac.uk/analysis/xrt/files/SWIFT-XRT-CALDB-09\_v20.pdf}, prepared by Beardmore and collaborators, and references therein.} and absorption model (\texttt{PHABS} was adopted) again do not play a major role in the $N_H(z)$ distribution.

\begin{figure}[tb]
	\centering
	\includegraphics[width=\columnwidth]{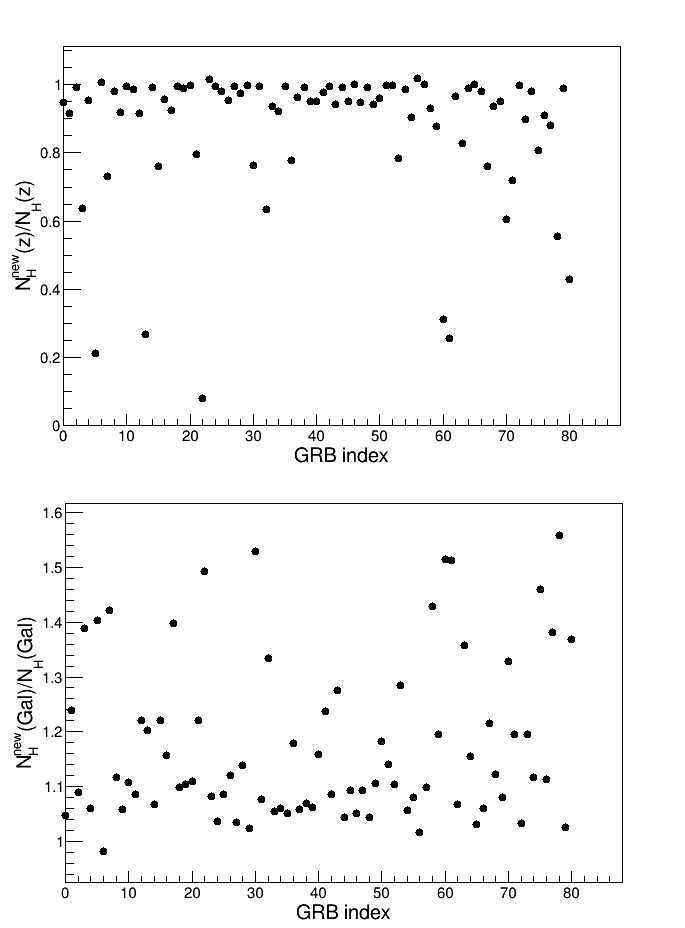}
	\caption{On the top panel it is reported the ratio between the new and the old intrinsic column densities, as a function of an index representing the 81 GRBs of Table~\ref{tab:longtable} with a measure of redshift. On the bottom panel, a comparable distribution of the ratio between the new (W13) and the old (LAB) Galactic column densities is shown.}
	\label{fig:RATIO}
\end{figure}

\begin{figure*}[tb]
	\centering
	\includegraphics[width=2.\columnwidth]{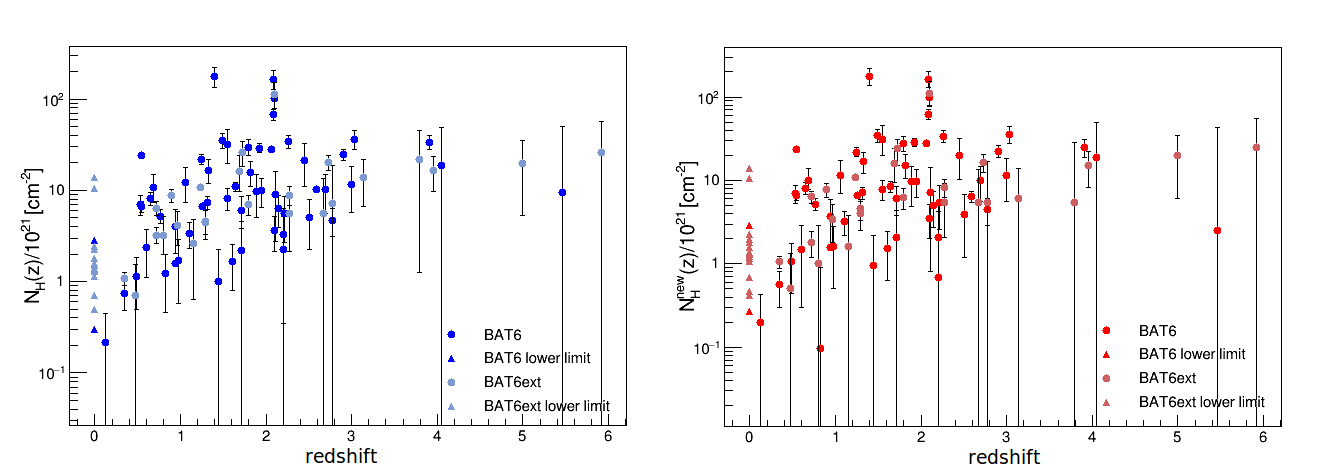}
	\caption{Intrinsic X-ray column density distribution as a function of redshift for both Galactic models. The panel on the left shows the $N_H(z)$ distribution obtained with the LAB survey. It is important to note that the 41 GRBs of the extended sample, reported in a lighter colour with respect to the (blue) dots of the BAT6 sample, do not affect the increasing trend that came out in C12. The panel on the right reports the $N^{new}_H(z)$ distribution, computed with the new W13 Galactic absorption model. The triangles, representing $N_H(z)$ for the 18 GRBs with no redshift measure, should be considered as lower limits, increasing with redshift as $\sim(1+z)^{2.4}$. The error bars were figured out within \texttt{XSPEC} at $90\%$ confidence level ($\Delta\chi^2=2.71$) and so were the upper limits.}
	\label{fig:distrib}
\end{figure*}

In Fig.~\ref{fig:RATIO}, the ratio between the new and the old intrinsic column densities is reported for all GRBs of Table~\ref{tab:longtable}. While it is clear that every new single value shows a decrease (with a maximum of a factor $\sim5$), most of them are only marginally affected by the different Galactic model assumed. This is due to the fact that, as we expected, the ratio between the new and the old Galactic column densities is, in almost all cases, just slightly greater than one (see bottom panel of Fig.~\ref{fig:RATIO}).

The redshift distributions of the X-ray intrinsic column density are shown in Fig.~\ref{fig:distrib}. The increasing trend of $N_H(z)$ with redshift is still evident, not only in the distribution obtained with the LAB Galactic model, but also in the distribution obtained with the new W13 Galactic model. This states that, even if the W13 model changes the single values, the whole $N_H(z)$ distribution follows the same increasing trend found in the previous works. Furthermore, note that the supplementary 41 GRBs of the extended BAT6 sample are distributed evenly among the other column densities, with no influence on the observed pattern. Hence, it seems that neither a radical improvement of the Galactic model nor an extension of the sample affect the observed general trend in the $N_H(z)$ distribution. This suggests that those should be considered as second-order corrections.
\subsection{Monte Carlo simulations and K-S test}
\begin{figure}[tb]
	\centering
	\includegraphics[width=\columnwidth]{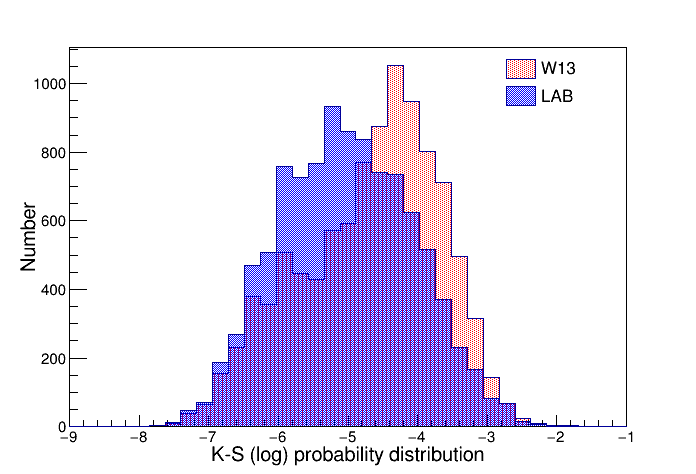}
	\caption{Kolmogorov-Smirnov (log) probability distributions. The dotted (red) histogram is related to the $N^{new}_H(z)$ values (obtained with W13 Galactic model), while the thicker dotted (blue) histogram is related to the probability distribution of $N_H(z)$ values (obtained with the LAB survey). In both cases, it is evident that the probability of having the two sub-samples descending from the same parent distribution is extremely low.}
	\label{fig:histo_probab}
\end{figure}

It seems that the uniform trend of absorbing column densities at low redshift does not persist at high redshift. In order to see if there is a real difference, so that the increasing trend could not be attributed to statistical fluctuations, we cut the sample at $z=1$ (where the distribution bends) and realised a Kolmogorov-Smirnov test on the two $N_H(z)$ sub-samples. Monte Carlo simulations have been used to provide 10,000 $N_H(z)$ values for each GRB, extracted from a Gaussian distribution peaked at the intrinsic column density values listed in Table~\ref{tab:longtable} (and with a width of $1\sigma$). Furthermore, for the 18 GRBs of the BAT6ext sample without a redshift measure, we simulated also $z$, extracting the values from the redshift probability density function obtained from the other 81 GRBs of the sample, and scaled the corresponding $N_H(z=0)$, previously simulated, with the factor $(1+z)^{2.4}$. For each iteration, we computed the K-S test and we obtained a distribution of (log) probabilities, with mean values $\log P=-5.02\pm0.95$ and $\log P=-4.68\pm0.99$, for the intrinsic column densities obtained with the LAB survey and with the W13 Galactic model, respectively. These values are extremely low, and, as we can see in Figure~\ref{fig:histo_probab}, even the tails of the two distributions never reach a $1\%$ probability.

In addition, a much more brutal technique has been used to include those 18 GRBs in the simulations. We fixed $z$ to be 1 and we simulated the $N_H(z)$ values, first attributing all the $z=1$ GRBs to the sub-sample of GRBs with redshift under $z=1$ and then to the other. Then, the K-S test has been carried out, obtaining in the former case mean values $\log P=-3.02\pm0.52$ and $\log P=-2.60\pm0.44$ (for the values obtained with the LAB survey and the W13 model, respectively), in the latter case mean values $\log P=-3.44\pm0.69$ and $\log P=-3.42\pm0.66$ (for LAB survey and W13 model, respectively). This technique was thought to be a useful proof, stating that even in the most extreme cases, that is allocating all the GRBs around the bending redshift, the difference between the two sub-distributions does not seem to be due to adverse statistics.

In all these cases, the probability of having the two sub-samples descending from the same parent distribution is extremely low. This states that there is a real difference between the two sub-distributions of column densities $N_H(z)$ in excess of the Galactic value. Therefore the increasing trend of the intrinsic column density as a function of redshift can not be due to statistical fluctuations. According to this test, one can not attribute all the absorption in excess to the host galaxy, leading to suppose that every single value of these $N_H(z)$ distributions actually includes a contribution that clearly increases with the redshift. One possibility to account for this is to invoke the overlooked component of the absorption along the IGM, that obviously increases with the distance of the GRB \citep{Behar:IGM,Campana:missing}. Alternatively, one has to postulate that the GRB environment gets denser with redshift.

Furthermore, we performed a K-S test in order to show that GRBs with an upper limit in the measure of $N_H(z)$ are uniformly distributed in redshift. This is important to assure that the increasing trend of $N_H(z)$ with the redshift is not due to an incapability of measuring low $N_H$ at high $z$. In our sample a measure of $N_H(z)$ is provided essentially for almost all GRBs, but the eventual presence of most of the upper limit measures at high redshift could give rise to some objections. Hence, we separated into two sub-samples the redshift of GRBs with a measure of $N_H(z)$ and the redshift of GRBs with an upper limit (denoted with a $\downarrow$ in Table~\ref{tab:longtable}) and computed a K-S test for each iteration of the Monte Carlo simulation. We obtained mean probabilities of $P=30\pm8\%$ and $P=25\pm9\%$ (for measures obtained with the LAB Galactic model and W13, respectively), indicating that the two sets of redshift originate from the same parent distribution.
\subsection{Correlations of $N_H(z)$}
We already highlighted the fact that the only selection effect of our sample is the peak flux limit and that it is also highly complete in redshift, providing the possibility to compare the rest frame physical properties of GRBs in an unbiased way. In this work, we analyzed the intrinsic X-ray absorbing column density redshift distribution. Moreover, the rest-frame peak energy $E^{rest}_p$ and the bolometric equivalent isotropic luminosity $L_{iso}$ have been computed by \citet{Pescalli:BAT6ext} for each GRB of the BAT6ext with a redshift measure. 

We now want to study the eventual correlation between $N_H(z)$ and these quantities\footnote{Note that in this work we considered the redshift of GRB 121123A to be zero, and that no energy or luminosity measure has been obtained for GRB 070306 in \citet{Pescalli:BAT6ext}. Hence, the following analysis involves 80 GRBs.}. However, given that the sample was selected above a peak flux threshold, $E^{rest}_p$ and, primarily, $L_{iso}$ have a tight correlation with redshift. This is due to the fact that, for a flux-limited sample, the only sources detected out to great distances are necessarily powerful \citep[see][]{Blundell:biasLz}. However, we do not expect any correlation between $N_H(z)$ and luminosities (or energies), but, since all these quantities correlate with $z$, we must take the precaution of doing a correlation test excluding the dependence on redshift. Indeed, the spurious correlation between $N_H(z)$ and $E^{rest}_p$ ($L_{iso}$) can be avoided by using a partial correlation analysis. First, we computed the Spearman rank correlation coefficient $r$ \citep{Spearman:rank} for each case, namely $r_{12}$, $r_{13}$ and $r_{23}$, where 1,2 represent $N_H(z)$ and $E^{rest}_p$ ($L_{iso}$), respectively, and 3 is $z$. Then, we obtained the correlation coefficient between 1 and 2, excluding the redshift bias:
\begin{equation}r_{12,3}=\frac{r_{12}-r_{13}r_{23}}{\sqrt{1-r_{13}^2}\sqrt{1-r_{23}^2}}\end{equation} (\citealp[see][chap.~27]{Kendall:stats}; \citealp{Padovani:stats}).

From the Spearman rank coefficient one can derive the associated probability value, to figure out the significance of the correlation. The coefficient between $N_H(z)$ ($N^{new}_H(z)$) and $E^{rest}_p$ is $r_{12,3}=0.056$ ($0.093$) and the related probability is $P=62\%$ ($41\%$), while the correlation coefficient between $N_H(z)$ ($N^{new}_H(z)$) and $L_{iso}$ is $r_{12,3}=0.052$ ($0.075$), with probability $P=65\%$ ($51\%$). Hence, the results of the partial correlation analysis state that $N_H(z)$ does not correlate with $E^{rest}_p$ and $L_{iso}$. Being $N_H(z)$ connected to the distribution of matter along the line of sight and $E^{rest}_p$ and $L_{iso}$ to the GRB prompt emission, this result is not surprising.

Furthermore, our data can be used to test the anticorrelation between the X-ray absorbing column density in excess of the Galactic value (evaluated in the observer rest frame) and redshift, pointed out by \citet{Grupe:anti}. In our work, $N_H(z=0)$ has been obtained, for the 81 GRBs of the sample with a redshift measure, dividing the intrinsic column density values listed in Table~\ref{tab:longtable} by the scaling factor $\sim(1+z)^{2.4}$. Then, we computed the Spearman rank correlation coefficient between $N_H(z=0)$ and $1+z$, obtaining $r=-0.28$ and $r=-0.26$, for the values obtained with the LAB survey and the W13 Galactic model, respectively. The related probabilities are $P=1\%$ and $P=1.8\%$, too ambiguous to draw any significant conclusion about the presence of a correlation. Hence, with our results we can not confirm that such a correlation exists.


%

\section{Conclusions}
In this work we focused on the eventual effect of a radical change in the Galactic absorption model on the $N_H(z)$ distribution. We derived the intrinsic absorbing column densities neglecting the absorption along the IGM and using two different Galactic absorption models, the Leiden Argentine Bonn $H_I$ survey and the most recent model provided by W13. The two $N_H(z)$ distributions have a mean value of $\log(N_H/\rm{cm}^{-2})=21.93\pm0.54$ and $\log(N_H/\rm{cm}^{-2})=21.84\pm0.61$, respectively. As we expected, the new single intrinsic column density values are smaller, even by a factor $\sim5$. Nonetheless, if on the one hand the new Galactic model considerably affects the single column density values, on the other hand there is no radical change in the distribution as a whole. Therefore, the contribution of Galactic column densities alone, no matter how improved, is not sufficient to change the observed general trend. To explain the cosmological increase of $N_H(z)$ as a function of redshift it is necessary to include the contribution of both the diffuse intergalactic medium and the intervening systems along the line of sight of the GRBs.

This study clearly leaves unanswered the question about the high column density in GRB spectra. Different studies \citep{Campana:complete,Campana:missing,Schady:gasionizz} show that the intrinsic column density of GRBs is high. A small contribution can come from the intervening matter \citep{Campana:missing}, but this is clearly not enough to account for the observed distribution. Several mechanisms have been envisaged involving photoionization of the circumburst medium (by the GRB itself or pre-existing) or dense H$_2$ regions \citep{Campana:2007,Campana:absFOIO,Campana:complete,Campana:missing,Watson:density,Watson:2013,Schady:gasionizz,Starling:evoluzIGM,Krongold:2013}. Our study demonstrates that the absorbing contribution of our Galaxy plays a minor role in that.
\begin{acknowledgements}
      This work made use of data supplied by the UK \emph{Swift} Science Data Centre at the University of Leicester. We thank the referee E. Behar for helpful comments. We also thank Paolo D'Avanzo for useful discussions.
\end{acknowledgements}

%
%
\bibliographystyle{aa} 
\bibliography{bibliografia1,bibliografia2} 

\onecolumn
\begin{longtab}
%

\begin{landscape}
\small
\begin{ThreePartTable}
  \begin{TableNotes}\footnotesize
	\item[a] This GRB has an inferred photometric redshift of $z=2.7$. Since this is not a spectroscopic measure in data analysis we considered $z$ to be zero, but for completeness we reported also the intrinsic column densities computed at $z=2.7$.
	\item[b] In this case the column density fitted was an upper limit (a potential $\downarrow$), but being relative to a GRB with no redshift, this should be also a lower limit. The symbol $\gtrless$ indicates this ambiguous situation.
  \end{TableNotes}
\LTcapwidth=1.1\textwidth
\begin{longtable}{cccccccccc}
\caption{Intrinsic column densities of the BAT6ext sample. A solid horizontal line separates BAT6 GRBs from the extended sample. GRBs with an updated redshift estimate with respect to C12 are marked in italics. The $N_H(z)$ values for the GRBs with no redshift are evaluated at $z=0$. The results with an upper limit of column density are labelled by $\downarrow$.}
\label{tab:longtable} \\
\toprule
\multicolumn{1}{c}{GRB}	& 
\multicolumn{1}{c}{$z$} 		& 
\multicolumn{1}{c}{$N_H(z)$} 	&
\multicolumn{1}{c}{$\Gamma$}		& 
\multicolumn{1}{c}{$N_H(Gal)$}& 
\multicolumn{1}{c}{Time-interval}&
\multicolumn{1}{c}{Obs. mode} &
\multicolumn{1}{c}{$N^{new}_H(z)$}& 
\multicolumn{1}{c}{$N^{new}_H(Gal)$}& 
\multicolumn{1}{c}{$\Gamma^{new}$}\\
 &
 &
\multicolumn{1}{c}{$(10^{21}\,\text{cm}^{-2})$}&
 &
\multicolumn{1}{c}{$(10^{20}\,\text{cm}^{-2})$}& 
\multicolumn{1}{c}{(s)}&
\multicolumn{1}{c}{(exposure time in ks)} &
\multicolumn{1}{c}{$(10^{21}\,\text{cm}^{-2})$}& 
\multicolumn{1}{c}{$(10^{20}\,\text{cm}^{-2})$}& 
 \\
\midrule
\endfirsthead
\multicolumn{3}{l}{\footnotesize Continued} \\
\toprule
\multicolumn{1}{c}{GRB}	& 
\multicolumn{1}{c}{$z$} 		& 
\multicolumn{1}{c}{$N_H(z)$} 		&
\multicolumn{1}{c}{$\Gamma$}		& 
\multicolumn{1}{c}{$N_H(Gal)$}& 
\multicolumn{1}{c}{Time-interval}&
\multicolumn{1}{c}{Obs. mode} &
\multicolumn{1}{c}{$N^{new}_H(z)$}& 
\multicolumn{1}{c}{$N^{new}_H(Gal)$}& 
\multicolumn{1}{c}{$\Gamma^{new}$}\\
 &
 &
\multicolumn{1}{c}{$(10^{21}\,\text{cm}^{-2})$}&
 &
\multicolumn{1}{c}{$(10^{20}\,\text{cm}^{-2})$}& 
\multicolumn{1}{c}{(s)}&
\multicolumn{1}{c}{(exposure time in ks)} &
\multicolumn{1}{c}{$(10^{21}\,\text{cm}^{-2})$}& 
\multicolumn{1}{c}{$(10^{20}\,\text{cm}^{-2})$}& 
 \\
\midrule
\endhead
\midrule
\insertTableNotes
\endfoot
\bottomrule
\insertTableNotes
\endlastfoot
$050318$ 	& $1.44$	& $\downarrow 1.00^{+1.24}_{-1.00}$ & $1.94\pm0.09$ & $1.87$ & $3000-7\times10^4$ & PC$(23.4)$ & $\downarrow 0.95^{+1.24}_{-0.95}$& $1.96$& $1.94\pm0.09$
\\
$050401$ 	& $2.90$	& $24.47^{+3.57}_{-3.41}$ & $1.86\pm0.05$ & $4.40$ & $200-2000$ & WT$(1.4)$ & $22.4^{+3.6}_{-3.4}$& $5.45$ & $1.86\pm0.05$
\\
$050416$A 	& $0.654$	& $8.07^{+1.40}_{-1.28}$ & $1.97\pm0.10$ & $2.45$ & $400-3\times10^5$ & PC$(92.3)$ & $8.0^{+1.4}_{-1.3}$& $2.67$ & $1.97\pm0.10$
\\
$050525$A 	& $0.606$	& $2.36^{+1.38}_{-1.25}$ & $1.97\pm0.16$ & $9.08$ & $6000-3\times10^5$ & PC$(18.7)$ & $1.5^{+1.4}_{-1.2}$& $12.6$ & $1.96\pm0.16$
\\
$050802$ 	& $1.71$	& $\downarrow 2.20^{+2.32}_{-2.20}$ & $1.80\pm0.11$ & $1.85$ & $400-2000$ & PC$(1.5)$ & $\downarrow 2.1^{+2.3}_{-2.1}$ & $1.96$ & $1.80\pm0.11$
\\
$050922$C 	& $2.198$	& $3.25^{+3.05}_{-2.90}$ & $2.12\pm0.10$ & $5.40$ & $400-3\times10^5$ & PC$(35.1)$ & $\downarrow 0.69^{+3.60}_{-0.69}$ & $7.58$ & $2.14\pm0.10$
\\
$060206$ 	& $4.048$	& $\downarrow 18.89^{+30.25}_{-18.89}$ & $2.00\pm0.31$ & $0.93$ & $700-10^5$ & PC$(4.5)$ & $\downarrow 19.0^{+30.0}_{-19.0}$ & $0.917$ & $1.99\pm0.30$
\\
$060210$ 	& $3.91$	& $33.77^{+6.20}_{-5.99}$ & $1.99\pm0.05$ & $6.08$ & $3000-10^6$ & PC$(76.8)$ & $24.7^{+6.3}_{-6.0}$ & $8.64$ & $2.00\pm0.05$
\\
$\emph{060306}$ 	& $1.55$ & $31.60^{+15.00}_{-11.66}$ & $1.77\pm0.26$ & $3.44$ & $300-925$ & PC$(0.6)$ & $31.0^{+15.0}_{-11.0}$ & $3.84$ & $1.77\pm0.25$
\\
$060614$ 	& $0.125$ & $\downarrow 0.22^{+0.23}_{-0.22}$ & $1.77\pm0.08$ & $1.88$ & $4400-10^6$ & PC$(204.3)$ & $\downarrow 0.20^{+0.23}_{-0.20}$ & $1.99$ & $1.77\pm0.07$
\\
$060814$ 	& $1.923$ & $28.93^{+3.41}_{-3.21}$ & $1.96\pm0.07$ & $2.33$ & $250-500$ & WT$(0.2)$ & $28.8^{+3.4}_{-3.2}$ & $2.58$ & $1.96\pm0.07$
\\
$060904$A 	& $-$ & $>2.84^{+0.79}_{-0.71}$ & $3.42\pm0.37$ & $1.33$ & $185-225$ & WT$(0.04)$ & $>2.85^{+0.79}_{-0.71}$ & $1.39$ & $3.42\pm0.37$
\\
$060908$ 	& $1.884$ & $9.73^{+5.28}_{-4.71}$ & $2.07\pm0.19$ & $2.35$ & $200-10^5$ & PC$(11.6)$ & $9.6^{+5.3}_{-4.7}$ & $2.55$ & $2.07\pm0.19$
\\
$060912$A 	& $0.94$ & $4.04^{+2.20}_{-2.00}$ & $1.92\pm0.18$ & $3.85$ & $200-2000$ & PC$(1.7)$ & $3.7^{+2.2}_{-2.0}$ & $4.70$ & $1.93\pm0.18$
\\
$060927$ 	& $5.467$ & $\downarrow 9.39^{+40.86}_{-9.39}$ & $1.85\pm0.20$ & $4.61$ & $100-10^4$ & PC$(3.5)$ & $\downarrow 2.5^{+40.8}_{-2.5}$ & $5.54$ & $1.86\pm0.20$
\\
$061007$ 	& $1.261$ & $6.71^{+0.35}_{-0.35}$ & $1.89\pm0.02$ & $1.77$ & $90-2000$ & WT$(1.9)$ & $6.65^{+0.35}_{-0.35}$ & $1.89$ & $1.89\pm0.02$
\\
$061021$ 	& $0.346$ & $0.74^{+0.26}_{-0.26}$ & $1.89\pm0.06$ & $4.53$ & $3000-3\times10^5$ & PC$(83.1)$ & $0.56^{+0.26}_{-0.26}$ & $5.53$ & $1.89\pm0.09$
\\
$061121$ 	& $1.314$ & $7.43^{+1.24}_{-1.18}$ & $1.80\pm0.06$ & $4.00$ & $600-3\times10^5$ & PC$(43.1)$ & $7.1^{+1.2}_{-1.2}$ & $4.63$ & $1.80\pm0.06$
\\
$061222$A 	& $2.088$ & $67.70^{+9.29}_{-8.68}$ & $1.99\pm0.09$ & $9.02$ & $3\times10^4-2\times10^5$ & PC$(49.5)$ & $65.5^{+9.2}_{-8.6}$ & $12.60$ & $2.00\pm0.09$
\\
$070306$ 	& $1.496$ & $34.99^{+6.62}_{-5.98}$ & $1.77\pm0.11$ & $2.85$ & $10^4-4\times10^4$ & PC$(6.0)$ & $34.8^{+6.6}_{-6.0}$ & $3.13$ & $1.77\pm0.11$
\\
$\emph{070328}$ 	& $2.063$ & $27.93^{+2.00}_{-1.94}$ & $2.05\pm0.04$ & $2.61$ & $350-1000$ & WT$(0.7)$ & $27.6^{+2.0}_{-1.9}$ & $2.88$ & $2.05\pm0.04$
\\
$\emph{070521}$ 	& $2.087$ & $164.51^{+40.15}_{-35.48}$ & $1.70\pm0.19$ & $2.92$ & $3000-10^4$ & PC$(1.9)$ & $164.0^{+40.0}_{-35.0}$ & $3.24$ & $1.70\pm0.19$
\\
$071020$ 	& $2.146$ & $6.27^{+2.42}_{-2.30}$ & $1.78\pm0.07$ & $5.12$ & $70-300$ & WT$(0.2)$ & $5.0^{+2.4}_{-2.3}$ & $6.25$ & $1.79\pm0.07$
\\
$071112$C 	& $0.82$ & $1.22^{+0.79}_{-0.76}$ & $1.73\pm0.08$ & $7.44$ & $90-280$ & WT$(0.2)$ & $\downarrow 0.1^{+0.8}_{-0.1}$ & $11.1$ & $1.75\pm0.08$
\\
$071117$ 	& $1.33$ & $16.54^{+5.39}_{-4.60}$ & $1.95\pm0.18$ & $2.33$ & $2900-6.2\times10^4$ & PC$(18.2)$ & $16.8^{+4.8}_{-4.6}$ & $2.52$ & $1.97\pm0.20$
\\
$080319$B 	& $0.937$ & $1.59^{+0.14}_{-0.14}$ & $1.73\pm0.02$ & $1.11$ & $800-2000$ & WT$(0.9)$ & $1.58^{+0.14}_{-0.14}$ & $1.15$ & $1.74\pm0.02$
\\
$080319$C 	& $1.95$ & $9.90^{+3.71}_{-3.42}$ & $1.56\pm0.10$ & $2.20$ & $200-3\times10^5$ & PC$(24.3)$ & $9.7^{+3.7}_{-3.4}$ & $2.39$ & $1.56\pm0.10$
\\
$080413$B 	& $1.10$ & $3.34^{+1.07}_{-1.02}$ & $1.89\pm0.08$ & $3.06$ & $6\times10^3-10^6$ & PC$(98.5)$ & $3.18^{+1.06}_{-1.00}$ & $3.43$ & $1.89\pm0.08$
\\
$080430$ 	& $0.767$ & $5.20^{+0.91}_{-0.86}$ & $1.96\pm0.09$ & $0.96$ & $5000-3\times10^5$ & PC$(44.6)$ & $5.18^{+0.91}_{-0.86}$ & $0.99$ & $1.96\pm0.09$
\\
$\emph{080602}$ 	& $1.82$ & $15.69^{+5.15}_{-4.61}$ & $1.91\pm0.14$ & $3.51$ & $200-800$ & PC$(0.6)$ & $15.3^{+5.1}_{-4.6}$ & $4.00$ & $1.91\pm0.14$
\\
$080603$B 	& $2.69$ & $10.12^{+4.81}_{-4.50}$ & $1.79\pm0.10$ & $1.24$ & $100-250$ & WT$(0.2)$ & $10.1^{+4.8}_{-4.5}$ & $1.27$ & $1.79\pm0.10$
\\
$080605$ 	& $1.64$ & $11.01^{+1.33}_{-1.27}$ & $1.66\pm0.04$ & $6.67$ & $100-800$ & WT$(0.6)$ & $8.4^{+1.3}_{-1.3}$ & $10.2$ & $1.67\pm0.04$
\\
$080607$ 	& $3.036$ & $36.07^{+8.79}_{-8.14}$ & $2.04\pm0.12$ & $1.69$ & $4000-6\times10^4$ & PC$(19.7)$ & $35.9^{+8.8}_{-8.2}$ & $1.82$ & $2.04\pm0.12$
\\
$080613$B 	& $-$ & $\gtrless0.3$\tnote{b} & $1.21\pm0.10$ & $3.17$ & $105-190$ & WT$(0.1)$ & $\gtrless0.27$\tnote{b} & $3.55$ & $1.20\pm0.11$
\\
$080721$ 	& $2.591$ & $10.11^{+0.92}_{-0.91}$ & $1.72\pm0.02$ & $6.95$ & $100-2000$ & WT$(1.3)$ & $6.40^{+0.93}_{-0.91}$ & $9.27$ & $1.73\pm0.02$
\\
$080804$ 	& $2.20$ & $\downarrow 2.24^{+2.89}_{-2.24}$ & $1.73\pm0.09$ & $1.63$ & $200-3\times10^5$ & PC$(41.0)$ & $\downarrow 2.1^{+2.9}_{-2.1}$ & $1.72$ & $1.73\pm0.09$
\\
$080916$A 	& $0.689$ & $10.73^{+4.16}_{-3.33}$ & $2.09\pm0.26$ & $1.84$ & $2\times10^4-10^6$ & PC$(110.2)$ & $9.9^{+4.0}_{-2.2}$ & $1.95$ & $2.04\pm0.18$
\\
$081007$ 	& $0.53$ & $6.94^{+1.81}_{-1.59}$ & $1.94\pm0.16$ & $1.37$ & $5000-4\times10^5$ & PC$(46.9)$ & $6.9^{+1.8}_{-1.6}$ & $1.44$ & $1.94\pm0.16$
\\
$081121$ 	& $2.512$ & $5.01^{+2.86}_{-2.74}$ & $1.85\pm0.07$ & $4.03$ & $3000-2\times10^6$ & PC$(128.0)$ & $3.9^{+2.8}_{-2.7}$ & $4.75$ & $1.85\pm0.07$
\\
$081203$A 	& $2.10$ & $3.63^{+1.54}_{-1.48}$ & $1.69\pm0.05$ & $1.71$ & $200-600$ & WT$(0.4)$ & $3.5^{+1.6}_{-1.5}$ & $1.81$ & $1.69\pm0.05$
\\
$081221$ 	& $2.26$ & $34.16^{+5.66}_{-5.26}$ & $1.96\pm0.09$ & $2.03$ & $300-500$ & WT$(0.2)$ & $33.9^{+5.6}_{-5.3}$ & $2.17$ & $1.96\pm0.09$
\\
$081222$ 	& $2.77$ & $4.74^{+1.62}_{-1.59}$ & $1.93\pm0.04$ & $2.23$ & $60-1000$ & WT$(0.8)$ & $4.5^{+1.6}_{-1.6}$ & $2.37$ & $1.93\pm0.04$
\\
$090102$ 	& $1.547$ & $8.20^{+2.29}_{-2.14}$ & $1.66\pm0.08$ & $4.05$ & $700-7\times10^4$ & PC$(19.3)$ & $7.8^{+2.3}_{-2.1}$ & $4.69$ & $1.67\pm0.08$
\\
$\emph{090201}$ 	& $2.1$ & $101.21^{+21.08}_{-21.93}$ & $1.81\pm0.14$ & $4.94$ & $3000-6000$ & WT$,(0.2)$ & $99.0^{+21.0}_{-19.0}$ & $6.11$ & $1.81\pm0.15$
\\
$090424$ 	& $0.544$ & $6.58^{+0.90}_{-0.84}$ & $1.95\pm0.08$ & $1.86$ & $2000-3\times10^6$ & PC$(213.1)$ & $6.55^{+0.89}_{-0.83}$ & $2.02$ & $1.95\pm0.08$
\\
$\emph{090709}A$ 	& $1.8$ & $29.45^{+6.20}_{-5.61}$ & $1.84\pm0.11$ & $6.62$ & $4000-1.5\times10^4$ & PC$(4.4)$ & $27.7^{+6.2}_{-5.6}$ & $8.44$ & $1.84\pm0.11$
\\
$090715$B 	& $3.00$ & $11.71^{+6.48}_{-6.02}$ & $1.92\pm0.12$ & $1.34$ & $4000-10^5$ & PC$(28.8)$ & $11.6^{+6.5}_{-6.0}$ & $1.40$ & $1.92\pm0.12$
\\
$090812$ 	& $2.452$ & $21.12^{+11.56}_{-9.97}$ & $2.08\pm0.24$ & $2.26$ & $10^4-7\times10^4$ & PC$(8.6)$ & $20.1^{+11.6}_{-9.9}$ & $2.47$ & $2.07\pm0.24$
\\
$090926$B 	& $1.24$ & $21.90^{+2.77}_{-2.60}$ & $1.88\pm0.08$ & $1.92$ & $130-300$ & WT$(0.2)$ & $21.9^{+2.8}_{-2.6}$ & $2.02$ & $1.88\pm0.08$
\\
$091018$ 	& $0.971$ & $1.69^{+1.21}_{-1.11}$ & $1.76\pm0.11$ & $2.81$ & $150-1000$ & PC$(0.9)$ & $1.6^{+1.2}_{-1.1}$ & $3.07$ & $1.76\pm0.11$
\\
$091020$ 	& $1.71$ & $6.05^{+2.42}_{-2.23}$ & $1.78\pm0.10$ & $1.39$ & $200-400$ & WT$(0.2)$ & $6.0^{+2.4}_{-2.2}$ & $1.45$ & $1.78\pm0.10$
\\
$091127$ 	& $0.49$ & $1.15^{+0.69}_{-0.65}$ & $1.77\pm0.12$ & $2.81$ & $6000-2\times10^4$ & PC$(1.6)$ & $1.08^{+0.69}_{-0.64}$ & $3.11$ & $1.74\pm0.11$
\\
$091208$B 	& $1.06$ & $12.10^{+5.90}_{-4.69}$ & $2.05\pm0.25$ & $4.86$ & $200-600$ & PC$(0.4)$ & $11.6^{+5.8}_{-4.6}$ & $5.75$ & $2.06\pm0.25$
\\
$\emph{100615}A$ 	& $1.398$ & $176.15^{+46.54}_{-41.09}$ & $2.25\pm0.34$ & $3.28$ & $200-2000$ & PC$(1.4)$ & $176.0^{+47.0}_{-40.0}$ & $3.74$ & $2.25\pm0.34$
\\
$100621$A 	& $0.542$ & $23.88^{+1.75}_{-1.17}$ & $2.67\pm0.09$ & $2.89$ & $130-200$ & WT$(0.1)$ & $23.8^{+1.7}_{-1.7}$ & $3.19$ & $2.68\pm0.09$
\\
$100728$B 	& $2.106$ & $9.05^{+7.18}_{-6.32}$ & $1.98\pm0.19$ & $6.17$ & $4000-3\times10^4$ & PC$(8.5)$ & $7.1^{+7.1}_{-6.3}$ & $7.93$ & $1.99\pm0.19$
\\
$110205$A 	& $2.22$ & $5.57^{+3.05}_{-2.88}$ & $2.02\pm0.11$ & $1.61$ & $5000-5\times10^4$ & PC$(18.8)$ & $5.5^{+3.1}_{-2.9}$ & $1.70$ & $2.02\pm0.11$
\\
$110503$A 	& $1.613$ & $1.67^{+0.90}_{-0.87}$ & $1.71\pm0.05$ & $2.63$ & $200-700$ & WT$(0.5)$ & $1.51^{+0.92}_{-0.87}$ & $2.84$ & $1.71\pm0.05$
\\
\midrule
$110709$A 	& $-$ & $>10.49^{+1.31}_{-1.22}$ & $1.92\pm0.12$ & $1.54$ & $600-9000$ & PC$(3.9)$ & $>10.7^{+1.3}_{-1.2}$ & $1.61$ & $1.93\pm0.13$
\\
$110709$B 	& $<4$ & $>1.50^{+0.20}_{-0.19}$ & $2.04\pm0.06$ & $5.58$ & $10^4-2\times10^5$ & PC$(45.6)$ & $>1.37^{+0.20}_{-0.19}$ & $6.91$ & $2.04\pm0.06$
\\
$110915$A 	& $-$ & $>2.43^{+1.06}_{-0.92}$ & $1.90\pm0.25$ & $1.61$ & $5000-8000$ & PC$(1.5)$ & $>1.95^{+1.08}_{-0.93}$ & $6.67$ & $1.90\pm0.25$
\\
$111008$A 	& $4.99$ & $19.63^{+15.40}_{-14.38}$ & $1.81\pm0.10$ & $0.98$ & $5000-3\times10^4$ & PC$(4.8)$ & $20.0^{+15.0}_{-14.0}$ & $0.992$ & $1.81\pm0.10$
\\
$111228$A 	& $0.714$ & $6.39^{+1.12}_{-1.03}$ & $2.05\pm0.09$ & $2.96$ & $4000-2\times10^6$ & PC$(144.8)$ & $6.4^{+1.0}_{-1.1}$ & $3.25$ & $2.05\pm0.08$
\\
$120102$A 	& $-$ & $>2.40^{+0.36}_{-0.34}$ & $1.91\pm0.08$ & $10.30$ & $4000-7\times10^4$ & PC$(18.1)$ & $>1.98^{+0.36}_{-0.34}$ & $14.7$ & $1.91\pm0.08$
\\
$120116$A 	& $-$ & $>1.32^{+0.75}_{-0.66}$ & $1.94\pm0.23$ & $4.64$ & $300-6000$ & PC$(1.5)$ & $>1.24^{+0.76}_{-0.66}$ & $5.53$ & $1.94\pm0.23$
\\
$120119$A 	& $1.728$ & $25.83^{+8.72}_{-7.50}$ & $1.65\pm0.13$ & $7.91$ & $7000-2\times10^5$ & PC$(18.9)$ & $24.0^{+7.3}_{-8.9}$ & $11.30$ & $1.67\pm0.13$
\\
$120326$A 	& $1.798$ & $7.07^{+1.79}_{-1.71}$ & $1.79\pm0.06$ & $5.24$ & $4000-10^5$ & PC$(21.4)$ & $6.2^{+1.8}_{-1.7}$ & $6.26$ & $1.80\pm0.06$
\\
$120703$A 	& $-$ & $>1.57^{+0.44}_{-0.41}$ & $1.82\pm0.14$ & $1.21$ & $3000-5\times10^4$ & PC$(9.8)$ & $>1.58^{+0.45}_{-0.41}$ & $1.27$ & $1.78\pm0.13$
\\
$120729$A 	& $0.8$ & $3.21^{+1.94}_{-1.22}$ & $1.68\pm0.13$ & $14.2$ & $300-7000$ & PC$(2.8)$ & $\downarrow 1.0^{+1.9}_{-1.0}$ & $21.5$ & $1.70\pm0.13$
\\
$120802$A 	& $3.796$ & $21.56^{+23.29}_{-20.29}$ & $1.83\pm0.17$ & $9.65$ & $100-1.5\times10^4$ & PC$(5.1)$ & $\downarrow 5.5^{+23.2}_{-5.5}$ & $14.6$ & $1.87\pm0.17$
\\
$120811$C 	& $2.671$ & $\downarrow 5.60^{+8.02}_{-5.60}$ & $1.58\pm0.15$ & $2.09$ & $200-1100$ & PC$(0.9)$ & $\downarrow 5.4^{+8.0}_{-5.4}$ & $2.23$ & $1.58\pm0.15$
\\
$120907$A 	& $0.97$ & $4.11^{+1.77}_{-1.62}$ & $1.81\pm0.13$ & $5.35$ & $280-8000$ & PC$(3.2)$ & $3.4^{+1.8}_{-1.6}$ & $7.26$ & $1.82\pm0.13$
\\
$\emph{121123}A$\tnote{a} & $-$ & $>0.49^{+0.30}_{-0.28}$ & $1.85\pm0.11$ & $4.01$ & $4500-42000$ & PC$(14.9)$ & $>0.42^{+0.30}_{-0.28}$ & $4.75$ & $1.85\pm0.11$
\\
 & $2.7$ & $8.64^{+5.11}_{-4.77}$ & $1.82\pm0.10$ &  & & & $7.37^{+5.12}_{-4.78}$ &  & $1.82\pm0.10$
\\
$121125$A 	& $-$ & $>1.15^{+0.33}_{-0.31}$ & $2.04\pm0.12$ & $1.27$ & $4000-3.5\times10^4$ & PC$(13.2)$ & $>1.15^{+0.33}_{-0.31}$ & $1.32$ & $2.04\pm0.12$
\\
$121209$A 	& $2.1$ & $112.15^{+40.69}_{-33.11}$ & $1.79\pm0.25$ & $3.81$ & $5000-7000$ & PC$(1.4)$ & $111.0^{+41.0}_{-33.0}$ & $4.40$ & $1.79\pm0.25$
\\
$130420$A 	& $1.297$ & $4.60^{+1.35}_{-1.28}$ & $2.16\pm0.10$ & $1.26$ & $1000-2\times10^5$ & PC$(35.5)$ & $4.6^{+1.3}_{-1.3}$ & $1.30$ & $2.16\pm0.10$
\\
$130427$A 	& $0.34$ & $1.08^{+0.18}_{-0.18}$ & $1.68\pm0.04$ & $1.80$ & $2\times10^4-2.5\times10^6$ & PC$(165.0)$ & $1.06^{+0.18}_{-0.18}$ & $1.91$ & $1.68\pm0.04$
\\
$130427$B 	& $2.78$ & $\downarrow 7.11^{+11.48}_{-7.11}$ & $1.71\pm0.19$ & $4.40$ & $200-2\times10^4$ & PC$(5.3)$ & $\downarrow 5.4^{+11.4}_{-5.4}$ & $5.35$ & $1.71\pm0.19$
\\
$130502$A 	& $-$ & $>1.80^{+0.98}_{-0.85}$ & $2.05\pm0.30$ & $2.49$ & $100-10^4$ & PC$(1.7)$ & $>1.79^{+0.99}_{-0.84}$ & $2.73$ & $2.05\pm0.30$
\\
$130505$A 	& $2.27$ & $8.86^{+2.08}_{-2.01}$ & $1.77\pm0.05$ & $3.59$ & $3000-10^5$ & PC$(16.3)$ & $8.3^{+2.1}_{-2.0}$ & $4.03$ & $1.77\pm0.05$
\\
$130527$A 	& $-$ & $>1.37^{+0.74}_{-0.65}$ & $1.46\pm0.17$ & $3.73$ & $450-2000$ & PC$(1.2)$ & $>1.32^{+0.75}_{-0.66}$ & $4.34$ & $1.46\pm0.17$
\\
$130606$A 	& $5.913$ & $\downarrow 26.28^{+30.77}_{-26.28}$ & $1.87\pm0.13$ & $1.98$ & $5000-6\times10^4$ & PC$(16.5)$ & $\downarrow 25.0^{+31.0}_{-25.0}$ & $2.14$ & $1.87\pm0.13$
\\
$130609$B 	& $-$ & $>1.30^{+0.37}_{-0.34}$ & $1.87\pm0.12$ & $1.45$ & $5000-8000$ & PC$(2.4)$ & $>1.30^{+0.37}_{-0.35}$ & $1.51$ & $1.87\pm0.12$
\\
$130701$A 	& $1.155$ & $2.64^{+2.20}_{-2.00}$ & $1.77\pm0.14$ & $6.93$ & $1850-3.5\times10^4$ & PC$(6.6)$ & $\downarrow 1.6^{+2.2}_{-1.6}$ & $9.20$ & $1.78\pm0.14$
\\
$130803$A 	& $-$ & $>13.83^{+3.07}_{-2.88}$ & $2.67\pm0.31$ & $4.15$ & $4000-6\times10^4$ & PC$(11.3)$ & $>14.0^{+3.0}_{-2.7}$ & $5.12$ & $2.68\pm0.31$
\\
$130831$A 	& $0.479$ & $\downarrow 0.71^{+0.82}_{-0.71}$ & $1.74\pm0.14$ & $4.80$ & $11000-1.2\times10^5$ & PC$(11.5)$ & $\downarrow 0.51^{+0.82}_{-0.51}$ & $5.74$ & $1.71\pm0.14$
\\
$130907$A 	& $1.238$ & $10.86^{+0.90}_{-0.87}$ & $1.82\pm0.04$ & $0.997$ & $7000-1.5\times10^5$ & PC$(37.3)$ & $10.84^{+0.92}_{-0.86}$ & $1.03$ & $1.82\pm0.04$
\\
$131030$A 	& $1.293$ & $4.45^{+1.62}_{-1.52}$ & $1.92\pm0.10$ & $4.70$ & $4000-90000$ & PC$(9.6)$ & $4.0^{+1.6}_{-1.5}$ & $5.62$ & $1.92\pm0.10$
\\
$131105$A 	& $1.686$ & $16.30^{+8.03}_{-6.66}$ & $1.83\pm0.21$ & $2.91$ & $350-1100$ & PC$(0.8)$ & $16.0^{+8.0}_{-6.6}$ & $3.25$ & $1.83\pm0.21$
\\
$140102$A 	& $-$ & $>0.70^{+0.22}_{-0.21}$ & $1.77\pm0.07$ & $2.71$ & $950-2300$ & WT$(1.3)$ & $>0.68^{+0.22}_{-0.21}$ & $3.04$ & $1.77\pm0.07$
\\
$140206$A 	& $2.73$ & $20.28^{+4.07}_{-3.89}$ & $1.81\pm0.06$ & $4.79$ & $10^4-10^5$ & PC$(22.7)$ & $16.4^{+4.1}_{-3.9}$ & $6.99$ & $1.83\pm0.06$
\\
$140215$A 	& $-$ & $>1.41^{+1.01}_{-0.88}$ & $1.65\pm0.23$ & $9.71$ & $300-700$ & PC$(0.4)$ & $>1.08^{+1.02}_{-0.88}$ & $13.1$ & $1.65\pm0.23$
\\
$140419$A 	& $3.956$ & $16.57^{+7.07}_{-6.75}$ & $1.83\pm0.07$ & $3.53$ & $1500-20000$ & PC$(7.8)$ & $15.1^{+7.1}_{-6.8}$ & $3.93$ & $1.84\pm0.07$
\\
$140506$A 	& $0.889$ & $8.74^{+1.58}_{-1.45}$ & $1.77\pm0.08$ & $7.67$ & $5000-10^5$ & PC$(18.9)$ & $7.7^{+1.6}_{-1.4}$ & $10.6$ & $1.78\pm0.08$
\\
$140512$A 	& $0.725$ & $3.24^{+0.64}_{-0.62}$ & $1.81\pm0.06$ & $9.43$ & $4500-7\times10^4$ & PC$(15.0)$ & $1.80^{+0.65}_{-0.61}$ & $14.7$ & $1.83\pm0.06$
\\
$140619$A 	& $-$ & $>2.24^{+0.54}_{-0.50}$ & $2.09\pm0.15$ & $1.58$ & $1700-25000$ & PC$(6.2)$ & $>2.25^{+0.54}_{-0.50}$ & $1.65$ & $2.09\pm0.15$
\\
$140628$A 	& $-$ & $>0.50^{+0.54}_{-0.47}$ & $1.86\pm0.20$ & $2.92$ & $750-2.2\times10^4$ & PC$(6.9)$ & $>0.47^{+0.55}_{-0.47}$ & $3.22$ & $1.86\pm0.20$
\\
$140629$A 	& $2.275$ & $5.56^{+3.67}_{-3.41}$ & $1.92\pm0.12$ & $0.909$ & $4000-1.5\times10^4$ & PC$(5.1)$ & $5.5^{+3.7}_{-3.4}$ & $0.932$ & $1.91\pm0.12$
\\
$140703$A 	& $3.14$ & $14.00^{+7.80}_{-7.27}$ & $1.80\pm0.09$ & $9.36$ & $3800-7\times10^4$ & PC$(11.9)$ & $\downarrow 6.0^{+7.8}_{-6.0}$ & $12.8$ & $1.82\pm0.09$
\\
\end{longtable}
\end{ThreePartTable}
\end{landscape}
\end{longtab}
\twocolumn

\end{document}